\documentclass[12pt]{article}
\usepackage{amsmath, amssymb, amsfonts, amsthm}
\title{Ground state energy of large atoms in a self-generated
magnetic field}
\author{L\'aszl\'o Erd\H os
 \thanks{Partially supported by SFB-TR12 of
the German Science Foundation}
\\Institute of Mathematics, University of Munich \\
Theresienstr. 39, D-80333 Munich, Germany
\\ and \\
Jan Philip Solovej \thanks{Work partially supported
   by the Danish Natural Science Research Council and by a Mercator
   Guest Professorship from the German Science Foundation}
\\ Department of Mathematics, University of Copenhagen\\
Universitetsparken 5, DK-2100 Copenhagen,\\
Denmark}

\date{\today}

\newtheorem{theorem}{Theorem}[section]

\newtheorem{lemma}[theorem]{Lemma}

\newcommand{\ul}{\underline}

\newcommand{\rd}{{\rm d}}
\newcommand{\be}{\begin{equation}}
\newcommand{\ee}{\end{equation}}
\newcommand{\bey}{\begin{eqnarray}}
\newcommand{\eey}{\end{eqnarray}}
\newcommand{\beys}{\begin{eqnarray*}}
\newcommand{\eeys}{\end{eqnarray*}}

\newcommand{\bB}{{\bf B}}
\newcommand{\bA}{{\bf A}}
\newcommand{\bZ}{{\bf Z}}
\newcommand{\bp}{{\bf p}}

\newcommand{\bsigma}{\mbox{\boldmath $\sigma$}}

\newcommand{\tchi}{\widetilde \chi}

\renewcommand{\iint}{\int \!\! \int}

\newcommand{\bR}{{\bf R}}
\newcommand{\bC}{{\bf C}}

\newcommand{\bP}{{\bf P}}

\newcommand{\bx}{{\bf x}}

\newcommand{\bN}{{\bf N}}

\newcommand{\ov}{\overline}

\newcommand{\e}{\varepsilon}

\newcommand{\Tr}{{\rm Tr\,}}

\newcommand{\wh}{\widehat}
\newcommand{\wt}{\widetilde}

\newcommand{\cF}{{\cal F}}
\newcommand{\cA}{{\cal A}}

\newcommand{\cE}{{\cal E}}

\newcommand{\cD}{{\cal D}}

\newcommand{\cN}{{\cal N}}

\newcommand{\cH}{{\cal H}}

\newcommand{\al}{\alpha}

\newcommand{\fh}{\frak h}

\begin{document}
\maketitle

\begin{abstract} We consider a large atom with nuclear charge $Z$
described by non-relativistic quantum mechanics with classical or
quantized electromagnetic field.  We prove that the absolute ground
state energy, allowing for minimizing over all possible
self-generated electromagnetic fields, 
is given by the non-magnetic Thomas-Fermi
theory to leading order in the simultaneous $Z\to \infty$, $\al\to 0$
limit if $Z\al^2\leq \kappa$ for some
universal $\kappa$, where $\al$ is the fine structure constant.
\end{abstract}

\bigskip\noindent
{\bf AMS 2000 Subject Classification: 81Q10}

\medskip\noindent
{\it Key words:} Pauli operator, magnetic field, quantum electrodynamics.

\medskip\noindent
{\it Running title:} Energy of atoms in a self-generated field


\section{Introduction}

The ground state energy of non-relativistic atoms and
molecules with large nuclear charge $Z$ can be described
by Thomas-Fermi theory to leading order in the $Z\to \infty$
limit \cite{L, LS}. Magnetic fields in this context were
taken into account only as an external field, either
a homogeneous one \cite{LSY1, LSY2}
or an inhomogeneous one \cite{ES} but subject to certain
regularity conditions. Self-generated magnetic fields, obtained
from Maxwell's equation are not known to satisfy these conditions.
In this paper we extend the validity of
Thomas-Fermi theory by allowing a self-generated magnetic field
that interacts with the electrons. This means, we look
for the absolute ground state of the system, after minimizing
for both the electron wave function and for the magnetic
field and we show that the additional magnetic field does
not change the leading order Thomas-Fermi energy.
Apart from finite energy, no other
 assumption is assumed on the magnetic field.

\bigskip

The nonrelativistic model of an atom in three spatial dimensions  with nuclear
charge $Z\ge 1$
and with $N$ electrons in a classical
magnetic field is given by
the Hamiltonian 
\be
   H_{N,Z}^{\rm cl}(\bA)=
 \sum_{j=1}^N \Big [T_j(\bA) - \frac{Z}{|x_j|}\Big]
  + \sum_{i<j} \frac{1}{|x_i-x_j|} + \frac{1}{8\pi\al^2}\int_{\bR^3} \bB^2
\label{hamcl}
\ee
acting on the space of antisymmetric functions $\bigwedge_1^N\cH$
with a single particle Hilbert space $\cH= L^2(\bR^3)\otimes \bC^2$.
The coordinates of the $N$ electrons are denoted by $\bx=
(x_1, x_2, \ldots, x_N)$.
The vector potential $\bA: \bR^3\to\bR^3$  generates
the magnetic field $\bB=\nabla\times \bA$ and it can be chosen
divergence-free, $\nabla\cdot \bA=0$. The last term in
\eqref{hamcl} is the energy of the magnetic field.
The kinetic energy 
of an electron is given by the Pauli operator
$$
   T(\bA) = [\bsigma\cdot (\bp+\bA)]^2= (\bp+\bA)^2 + \bsigma\cdot\bB, \qquad
  \bp= -i\nabla_x\, .
$$
Here $\bsigma$ is the vector of Pauli matrices.
We use the convention that for any one-body operator $T$, the
subscript in $T_j$ indicates that the operator acts on the $j$-th
variable, i.e. $T_j(\bA)= [\bsigma_j\cdot (-i\nabla_{x_j}+ \bA(x_j))]^2$.
The term $-Z|x_j|^{-1}$ describes the  attraction 
of the $j$-th electron to the nucleus located at the origin
and the term $|x_i-x_j|^{-1}$ is the electrostatic repulsion
between the $i$-th and $j$-th electron.

Our units are $\hbar^2(2me^2)^{-1}$ for the length, $2me^4\hbar^{-2}$
for the energy and $2mec\hbar^{-1}$ for the magnetic vector
potential, where $m$ is the electron mass, $e$ is the electron charge and
$\hbar$ is the Planck constant. 
 In these units, the only physical parameter
that appears in \eqref{hamcl} 
is the dimensionless fine structure constant $\al=e^2(\hbar c)^{-1}$.
We will assume that $Z\al^2\leq\kappa$ with some sufficiently small
universal constant $\kappa\le 1$
and we will investigate the simultaneous limit $Z\to \infty$, $\al\to 0$.
 Note that the field energy is
added to the total energy of the system and by the condition $\nabla\cdot \bA
=0$ we have
\be
\int_{\bR^3} \bB^2 = \int_{\bR^3} |\nabla\otimes \bA|^2,
\label{nablacurl}
\ee
where $\nabla\otimes \bA$ denotes the $3\times 3$ matrix
of all derivatives $\partial_iA_j$
and $|\nabla\otimes \bA|^2 =\sum_{i,j=1}^3|\partial_iA_j|^2$.
 We will always assume
that the vector potential belongs to the space of divergence free 
$H^1$-vector fields
$$
 \cA:= \{ \bA\in H^1(\bR^3, \bR^3), \; \nabla\cdot \bA=0\}.
$$

In the analogous nonrelativistic model of quantum electrodynamics,
the electromagnetic vector potential is quantized. In the
Coulomb gauge it is given by
$$
\bA_\Lambda(x) = \bA(x)= \bA_-(x) + \bA_-(x)^*
$$
with
$$
  \bA_-(x) = \frac{\al^{1/2}}{2\pi}\int_{\bR^3} \frac{g(k)}{\sqrt{|k|}}
  \sum_{\lambda=\pm} a_\lambda(k) {\bf e}_\lambda(k) e^{ik\cdot x} \rd k.
$$
Here $g(k)$ is a cutoff function, satisfying $|g(k)|\leq 1$ and
$\mbox{supp}\, g \subset \{k\in \bR^3\; : \; |k|\leq \Lambda\}$
with a constant $\Lambda<\infty$ (ultraviolet cutoff).
The field operators $\bA(x)$ depend on the cutoff function $g(k)$ whose
precise form is unimportant
the only relevant   
parameter is $\Lambda$.
For each $k$, the two polarization vectors ${\bf e}_-(k), {\bf e}_+(k)\in 
\bR^3$
are chosen such that together with the direction of propagation $k/|k|$
they are orthonormal. The operators $a_\lambda(k)$, $a_\lambda(k)^*$
are annihilation and creation operators acting on the bosonic
Fock space $\cF$ over $L^2(\bR^3)$ and satisfying the
canonical commutation relations
$$
[a_\lambda(k), a_{\lambda'}(k')]=
[a_\lambda(k)^*, a_{\lambda'}(k')^*]=0, \quad
[a_\lambda(k), a_{\lambda'}(k')^*]=\delta_{\lambda\lambda'}\delta(k-k').
$$
The field energy is given by
$$
   H_f = \al^{-1}\int_{\bR^3} |k| \sum_{\lambda= \pm}
  a_\lambda(k)^*a_\lambda(k) \rd k.
$$
The total Hamiltonian  is
\be
   H_{N,Z}^{\rm qed}= \sum_{j=1}^N \Big [T_j(\bA_\Lambda)
 - \frac{Z}{|x_j|}\Big]
  + \sum_{i<j} \frac{1}{|x_i-x_j|} + H_f
\label{hamqed}
\ee
and it acts on $(\bigwedge_1^N \cH)\otimes \cF$.

\bigskip

The stability of atoms in a classical
magnetic field \cite{F, FLL,LL,LLS} implies that
the operator \eqref{hamcl} is bounded from below
uniformly in $\bA$. If $Z\al^2$ is small enough.
It is known \cite{LY, ES2} that stability fails if $Z\al^2$ is too large.
 The analogous stability result
for quantized field \cite{BFG} states that \eqref{hamqed}
is bounded from below if $Z\al^2$ is small. 
In particular,  we can for
each fixed $\bA$ define the operators in (\ref{hamcl}) and
(\ref{hamqed}) as the
Friedrichs extensions of these operators defined on smooth functions
with compact support. 

The ground state energy of the operator with a classical field is given by
$$
   E_{N,Z}^{\rm cl}(\bA)
 = \inf\Big\{ \langle\Psi, H_{N,Z}^{\rm cl}(\bA)\Psi\rangle \; :
\;  \Psi\in \bigwedge_1^N \big(C_0^\infty(\bR^3)\otimes \bC^2\big),
 \; \|\Psi\|=1\Big\},
$$
and after minimizing in $\bA$ we set
$$
  E_{N,Z}^{\rm cl} = \inf_{\bA\in \cA}  E_{N,Z}^{\rm cl}(\bA).
$$
We note that it is sufficient to minimize over all $\bA\in \cA_0$ where
$\cA_0=H^1_c(\bR^3, \bR^3)$ denotes the space of compactly supported
$H^1$ vector fields. 
It is easy to see that the Euler-Lagrange equations for the
  above minimizations in $\Psi$ and $\bA$ correspond to the stationary
  version of the coupled
  Maxwell-Pauli system, i.e., the eigenvalue problem $H_{N,Z}^{\rm
    cl}(\bA)\Psi=E_{N,Z}^{\rm cl} \Psi$ together with the Maxwell equation
  $\nabla\times\bB=4\pi\alpha^2 {\bf
    J}_\Psi$, where ${\bf J}_\Psi$ is the current of the wave function
  $\Psi$. It is for this reason that it is natural to refer to $\bB$ as
a self-generated magnetic field in this context.

In the case of the quantized field, we define
$$
   E_{N,Z}^{\rm qed}
 = \inf\Big\{ \langle\Psi, H_{N,Z}^{\rm qed}\Psi\rangle \; :
\;  \Psi\in \bigwedge_1^N \big(C_0^\infty(\bR^3)\otimes \bC^2\big) \otimes
\cF,\; \|\Psi\|=1\Big\}.
$$
The stability results of \cite{F, FLL,LL,LLS, BFG} 
imply that $E_{N,Z}^{\rm cl}>-\infty$ and
 $E_{N,Z}^{\rm qed}>-\infty$  if $Z\al^2$ is small enough.

Finally, we define the ground state energy with no magnetic field as
$$
    E_{N,Z}^{\rm nf}:=  E_{N,Z}^{\rm cl}(\bA=0).
$$
In all three cases we define
$$
  E_Z^\#: = \inf_{N\in \bN}
 E_{N,Z}^\# , \qquad \#\in\{ {\rm cl}, {\rm qed}, {\rm nf}\},
$$
for the absolute (grand canonical) ground state energy.
The main result of this paper  states that the magnetic field does
not change the leading term of the absolute ground state energy
of a large atom in the $Z\to \infty$ limit. In particular,
Thomas-Fermi theory is correct to leading order even
with including self-generated magnetic field.

\begin{theorem}\label{thm:main1}
There exists a positive constant $\kappa$ such that 
if $Z\al^2\leq\kappa$, then
\be
   E_Z^{\rm nf}\ge 
 E_Z^{\rm cl} \ge E_Z^{\rm nf} - CZ^{\frac{7}{3}-\frac{1}{63}}.
\label{maincl}
\ee
For the quantized case, if we additionally 
assume 
\be
\Lambda\le \kappa^{\frac{1}{4}} Z^{\frac{7}{12}-\gamma}
\al^{-\frac{1}{4}}
\label{Lam}
\ee
 with some 
$0\leq\gamma\leq \frac{1}{63}$, then
\be
   E_Z^{\rm nf} + CZ\al\Lambda^2\ge 
 E_Z^{\rm qed} \ge E_Z^{\rm nf} - CZ^{\frac{7}{3}-\gamma}.
\label{mainqed}
\ee
We note that  $Z\al \Lambda^2 \ll Z^{7/3}$ 
if $\Lambda\ll \kappa^{-1/4} Z^{11/12}$ in the $Z\to\infty$ limit.
\end{theorem}

{\it Remark 1.} The leading term asymptotics of the non-magnetic
problem is given by the Thomas-Fermi theory and $ E_Z^{\rm nf}=-
c_{{\rm TF}} Z^{\frac{7}{3}} + O(Z^2)$ as $Z\to\infty$, where $c_{{\rm
    TF}}=3.678 \cdot (3\pi^2)^{\frac{2}{3}}$ is the Thomas-Fermi
constant. The leading order asymptotics was established in \cite{LS}
(see also \cite{L}). The correction to order $Z^2$ is known as the
Scott correction and was established in \cite{H,SW1} and for molecules
in \cite{IS} (See also \cite{SW2,SW3,SS}). The next term in the
expansion of order $Z^{5/3}$ was rigorously established for atoms in
\cite{FS}.
\label{rem:Scott}

\medskip

{\it Remark 2.} The exponents in the error terms are far from
being optimal. They can be improved by strengthening
our general semiclassical result Lemma \ref{lm:sc} for special
Coulomb like potentials using multi-scale analysis.

\medskip

{\it Remark 3.} For simplicity, we state and prove our results for
atoms, but the same proofs work
for molecules as well, if the number of nuclei $K$ is fixed, each has
a charge $Z$,  and
assume that the nuclei centers $\{ R_1, \ldots, R_K\}$
 are at least at distance $Z^{-1/3}$
away, i.e. $|R_i-R_j|\ge cZ^{-1/3}$, $i\neq j$.

\medskip

{\it Remark 4.} Theorem \ref{thm:main1} holds for the magnetic Schr\"odinger
operator as well, i.e. if we replace the Pauli
operator $T(\bA)= [\bsigma\cdot (\bp +\bA)]^2$ 
by $T(\bA)= (\bp +\bA)^2$ everywhere.
The proof is a trivial modification
of the Pauli case. The argument is in fact even easier;
instead of the magnetic Lieb-Thirring
inequality for the Pauli operator
 one uses the usual Lieb-Thirring inequality
that holds for the magnetic Schr\"odinger operator
uniformly in the magnetic field. We leave the
details to the reader.
Note that although the condition $Z\al^2\le \kappa$ is not needed in
order to ensure stability in the Schr\"odinger case, we still need it
  in the statement in Theorem \ref{thm:main1}. In the Schr\"odinger
  case this condition is not optimal.

\medskip
The upper bound in \eqref{maincl} is trivial by using a non-magnetic
trial state. The upper bound in \eqref{mainqed}
is obtained by a trial state that is
the tensor product of a non-magnetic electronic trial function
with the vacuum $|\Omega\rangle$ of $\cF$. 
The field energy $H_f$ and all terms that are linear
in $\bA$ give zero expectation value in the vacuum. The only
effect of the quantized field is in the nonlinear term $\bA^2$.
A simple calculation shows that
$$
  \langle \Omega| \bA^2|\Omega\rangle \leq C\al \Lambda^2.
$$

The main task is to prove the
lower bounds. Using the results from \cite{BFG},
the result for the quantized field \eqref{mainqed}
will directly follow from an analogous result
for a slightly modified Hamiltonian with a classical field.
Let
\be
   H_{N,Z}(\bA)= H_{N,Z,\al}(\bA)=
\sum_{i=1}^N \Big [T_i(\bA) - \frac{Z}{|x_i|}\Big]
  + \sum_{i<j} \frac{1}{|x_i-x_j|} + \frac{1}{8\pi \al^2}\int_{|x|\leq 3 r} 
|\nabla\otimes\bA|^2
\label{ham}
\ee
with some
\be
r= DZ^{-1/3} \;\;\mbox{with}\;\; D\ge 1.
\label{rchoice}
\ee
Note that instead of the local field energy, the total local $H^1$-norm
of $\bA$ is added in \eqref{ham}. By \eqref{nablacurl}, we have
\be
   H_{N,Z}^{\rm cl}(\bA)\ge H_{N,Z}(\bA)
\label{trivb}
\ee
for any $\bA\in \cA$. We define the ground state energy
of the modified Hamiltonian \eqref{ham}
$$
   E_{N,Z}(\bA):
 = \inf\Big\{ \langle\Psi, H_{N,Z}(\bA)\Psi\rangle \; :
\;  \Psi\in \bigwedge_1^N \big(C_0^\infty(\bR^3)\otimes \bC^2\big),
 \; \|\Psi\|=1\Big\}, 
$$
and set
$$
   E_{N,Z} := \inf_{\bA\in \cA}  E_{N,Z}(\bA), \quad
  E_Z : = \inf_N E_{N,Z},
$$
where the infimum for  $\bA\in \cA$ can again be restricted
to compactly supported vector potentials $\bA\in \cA_0$.  
For the modified classical Hamiltonian we have the
following theorem:

\begin{theorem}\label{thm:main} Let $Z\al^2\leq \kappa$ and
assume that $r= DZ^{-1/3}$ with $1\leq D\le Z^{1/63}$. Then
\be
   E_Z^{\rm nf}\ge  E_Z \ge E_Z^{\rm nf} - CZ^{7/3}D^{-1}.
\label{eq:main}
\ee
\end{theorem}

Taking into account \eqref{trivb}, Theorem \ref{thm:main}
 immediately implies the lower bound in \eqref{maincl}. The proof of
the lower bound in \eqref{mainqed} follows from  Theorem \ref{thm:main}
adapting an argument in \cite{BFG} that we will 
review in Section \ref{sec:q} for completeness.

One of the key ingredients of the proof of Theorem \ref{thm:main}
is  the following semiclassical statement
 that is of interest in itself. The first version is formulated under
general conditions but without an effective error term. In our
proof we actually use the second version that has 
a quantitative error term.

\begin{theorem}\label{lm:sc} 
Let $T_h(\bA)= [\bsigma\cdot(h\bp + \bA)]^2$ or 
$T_h(\bA)= (h\bp + \bA)^2$, 
$h\le 1$, and $V\ge 0$.

1) If $V\in L^{5/2}(\bR^3)\cap L^4(\bR^3)$,  then
\be
   \Tr \big[ T_h(\bA)-V\big]_- + h^{-2}\int_{\bR^3} \bB^2 \ge 
   \Tr \big[ -h^2\Delta - V\big]_- + o(h^{-3}) \quad \mbox{as $h\to0$}.
\label{sc1}
\ee

2) Assume that  $\| V\|_\infty\leq K$ with some $1\leq K \leq Ch^{-2}$
and consider the operators
with Dirichlet boundary condition on $\Omega\subset \bR^3$. 
Let $B_R$ denote the ball of radius $R$ about the origin and
let $\Omega_{\sqrt{h}} 
:= \Omega + B_{\sqrt{h}}$ denote the $\sqrt{h}$-neighborhood
of the set $\Omega$. We set $|\Omega_{\sqrt{h}}|$ for the
Lebesgue measure of $\Omega_{\sqrt{h}}$.
 Then
\be
\begin{split}
   \Tr \big[ (T_h(\bA)& -V)_{\Omega}\big]_- + h^{-2}\int_{\bR^3} \bB^2 \cr
&\ge
   \Tr \big[ (-h^2\Delta - V)_{\Omega}\big]_- 
 - Ch^{-3}K^{5/2}|\Omega_{\sqrt{h}}|
\big(h K^{3/2}\big)^{1/2} 
\big[1+\big(h K^{3/2}\big)^{1/2}  \big].
\label{sc3}
\end{split}
\ee
\end{theorem}

\noindent
{\it Remark.} Despite that the electrons are confined to $\Omega$,
their motion generates a magnetic field in the whole
$\bR^3$, so the magnetic field energy in \eqref{sc3}
is given by integration over  $\bR^3$.

\bigskip

We use the convention that letters $C, c$ denote positive
universal constants whose values may change from line to line.

\section{Reduction to the Main Lemmas}

{\it Proof of Theorem \ref{thm:main}.} We focus on the
lower bound, the upper bound is trivial. We start with
two localizations, one on scale $r\ge Z^{-1/3}$ and  the
other one on scale $d\le Z^{-1/3}$. The first one 
is designed to address
the difficulty that the $H^1$-norm of $\bA$
is available only locally around the nucleus.
This step would not be needed for the direct proof 
of \eqref{maincl}.
The second localization 
removes the ``Coulomb tooth'', i.e. the Coulomb singularity
near the nucleus.

In this section we reduce the proof of the lower bound in
Theorem \ref{thm:main}
to two lemmas. Lemma \ref{lm:h1} will show that the
Coulomb tooth is indeed negligible. Lemma \ref{lm:removefield}
shows that the magnetic field cannot substantially
lower the energy for the problem without the Coulomb tooth.
In the proof of Lemma \ref{lm:removefield}
we will use Theorem \ref{lm:sc}.

\bigskip

Recall that $B_R$ denotes the ball of radius $R$ about the origin.
We construct a pair of smooth cutoff functions satisfying 
the following conditions
$$
  \theta_0^2 + \theta_1^2\equiv 1, \;\; \mbox{supp}\; \theta_1 \subset B_{2d},
\quad \theta_1\equiv 1 \;\;\mbox{on $B_d$}, 
\quad |\nabla \theta_0|, \; |\nabla \theta_1|\leq Cd^{-1}.
$$
We will choose
\be
       d = \delta Z^{-1/3}
\label{deltachoice}
\ee
with some $\delta \le 1$, in particular $d\le r$.

\bigskip

We split the Hamiltonian as
$$
  H_{N,Z}(\bA) = H^0_{N,Z}(\bA) + H^1_{N,Z}(\bA)
$$
with
\be
\begin{split}\label{h0h1}
   H_{N,Z}^0(\bA)= & \sum_{i=1}^N 
 \Bigg[\theta_0\Big(T(\bA) -
 \frac{Z}{|x|} - \big(|\nabla\theta_0|^2
 +|\nabla\theta_1|^2\big)\Big)\theta_0\Bigg]_i\cr
 & + \sum_{i<j} \frac{1}{|x_i-x_j|} + \frac{1}{16\pi \al^2}\int_{B_{3r}} 
|\nabla\otimes\bA|^2\cr
   H_{N,Z}^1(\bA)=& \sum_{i=1}^N \Bigg[ 
\theta_1\Big (T(\bA) -
\frac{Z}{|x| }- \big(|\nabla\theta_0|^2
 +|\nabla\theta_1|^2\big)\Big)\theta_1\Bigg]_i
  +  \frac{1}{16 \pi\al^2}\int_{B_{3r}}|\nabla\otimes \bA|^2,
\end{split}
\ee
where we used the IMS localization formula that is valid for the
Pauli operator as well as for the Schr\"odinger operator.

In Section \ref{sec:tooth}
 we deal with $H_{N,Z}^1$, to prove that it is negligible:

\begin{lemma}\label{lm:h1} There is a positive universal constant $\kappa$ such
that for any $Z, \al$ with $Z\al^2\leq \kappa$ we have
$$
   \inf_N \inf_{\bA\in \cA_0}
 H_{N,Z}^1 (\bA) \ge - CZ^{7/3}\delta^{1/2} - Z^{2/3}
\delta^{-2}
$$
if $C Z^{-2/3}\le \delta\le D$ with a sufficiently large constant $C$.
\end{lemma}

Starting Section \ref{sec:h0} we will treat $H_{N,Z}^0(\bA)$ and 
we prove the following

\begin{lemma}\label{lm:removefield}
 There is a positive universal constant $\kappa$ such
that for any $Z, \al$ with $Z\al^2\leq \kappa$ we have
\be
    \inf_N \inf_\bA H_{N,Z}^0 (\bA) \ge
- c_{{\rm TF}} Z^{7/3} -C Z^{7/3}\big[
 Z^{-1/30} +D^{-1}\big]
\label{eq:remove}
\ee
with a sufficiently large constant $C$
if $ Z^{-1/6}\leq \delta\leq 1$
 and $D \le Z^{1/24}\delta^{13/16}$.
\end{lemma}
\noindent
The main ingredient in the proof 
is Theorem \ref{lm:sc} that will be proven in
Section \ref{sec:sc}. 
The proof of the lower bound in Theorem \ref{thm:main} then follows
from Lemmas \ref{lm:h1} and \ref{lm:removefield}
after choosing $\delta = Z^{-2/63}$.
$\Box$

\section{Estimating the Coulomb Tooth}\label{sec:tooth}

{\it Proof of the Lemma \ref{lm:h1}.}
Let $\tchi_0$ be a smooth cutoff function supported on $B_{3r}$
such that $|\nabla \tchi_0|\leq Cr^{-1}$ and $\tchi_0\equiv 1$
on $B_{2r}$. Let 
$ \langle \bA\rangle:=|B_{3r}|^{-1}\int_{B_{3r}} \bA$.
We define 
\be
   \bA_0:= (\bA-\langle\bA\rangle) \tchi_0, \qquad \bB_0:=\nabla\times \bA_0,
\label{a0}
\ee
then $\nabla \otimes\bA_0 = \tchi_0 \nabla\otimes \bA
+ (\bA-\langle\bA\rangle)\otimes\nabla\tchi_0$. Clearly
\be
\begin{split}
  \int_{\bR^3} \bB_0^2\leq \int_{\bR^3} |\nabla\otimes\bA_0|^2
 & \leq 2\int_{\bR^3} \tchi_0^2 |\nabla\otimes \bA|^2 + Cr^{-2}\int_{B_{3r}} 
(\bA-\langle\bA\rangle)^2  \cr
& \leq  C_1\int_{B_{3r}} |\nabla\otimes\bA|^2
\end{split}
\label{bb0}
\ee
for some universal constant $C_1$, 
where in the last step we used the Poincar\'e inequality.
Let
 $\varphi$ be a real phase  such that  $\nabla\varphi=\langle \bA\rangle$.
Since $\tchi_0\equiv 1$ on the support of $\theta_1$ by $D\ge \delta$, we have
$$
   \theta_1 T(\bA) \theta_1= \theta_1  e^{-i\varphi}
T(\bA-\langle\bA\rangle) e^{i\varphi}\theta_1 =
\theta_1  e^{-i\varphi} T(\bA_0) e^{i\varphi}\theta_1 .
$$
After these localizations, we have
\be
\begin{split}\label{h1}
   H_{N,Z}^1(\bA)\ge  &\sum_{j=1}^N
\Bigg[ \theta_1e^{-i\varphi}\Big (T(\bA_0) -
W(x)\Big)e^{i\varphi}\theta_1\Bigg]_j
+\frac{1}{2C_1\al^2}\int \bB_0^2
\end{split}
\ee
with
$$
   W(x) =\Big[\frac{Z}{|x|} + Cd^{-2}\Big]{\bf 1}(|x|\leq 2d)\, .
$$

Now we use the ``running energy scale''
argument in \cite{LLS}.
\be
\begin{split}
    \sum_{j=1}^N \Bigg[\theta_1
e^{-i\varphi}\Big [T(\bA_0)&- W\Big]e^{i\varphi}\theta_1\Bigg]_j
  \ge -\int_0^\infty\cN_{-e}( T(\bA_0)- W)\rd e\cr
 &\ge -\int_0^\mu\cN_{-e}(T(\bA_0)- W)\rd e -
\int_\mu^\infty\cN_{0}\Big(\frac{\mu}{e}T(\bA_0)- W+e\Big)\rd e\cr
 &\ge -\int_0^\mu\cN_{-e}(T(\bA_0)- W)\rd e -
\int_\mu^\infty\cN_{0}\Big(T(\bA_0)-\frac{e}{\mu} W+
\frac{e^2}{\mu}\Big)\rd e,
\end{split}\label{running}
\ee
where $\cN_{-e}(A)$ denotes the number of eigenvalues of a
self-adjoint operator $A$ below $-e$.

In the first term we use the bound $T(\bA_0)\ge (\bp+\bA_0)^2 - |\bB_0|$
and the CLR bound:
\be
\begin{split}
  \int_0^\mu\cN_{-e}(T(\bA_0)- W)\rd e &\leq C
  \int_0^\mu \rd e\int_{\bR^3}
    (W+ |\bB_0|-e)_+^{3/2}\cr
 & \le C\int_0^\mu \rd e\int_{\bR^3}
    (W-e/2)_+^{3/2} +  C\int_0^\mu \rd e\int_{\bR^3}
    (|\bB_0|-e^2/2\mu)_+^{3/2}\cr
&\le C\int_{\bR^3} W^{5/2} + C\mu^{1/2}\int_{\bR^3} \bB^2_0\cr
& = CZ^{5/2}d^{1/2}+Cd^{-2} + C\mu^{1/2}\int_{\bR^3} \bB^2_0.
\label{firstclr}
\end{split}
\ee
In the second term of \eqref{running} we use 
$$
  T(\bA_0)-\frac{e}{\mu} W \ge  \frac{1}{2}\Big[ (\bp +\bA_0)^2
 - \frac{2eZ}{\mu|x|}{\bf 1} (|x|\leq 2 d)\Big] + 
 \frac{1}{2} (\bp +\bA_0)^2 
 - |\bB_0|  -\frac{Ce}{\mu d^2}{\bf 1}(|x|\leq 2d),
$$
and that
\be
    (\bp +\bA_0)^2  - \frac{2eZ}{\mu|x|}{\bf 1} (|x|\leq 2d) \ge
(\bp +\bA_0)^2  - \frac{4eZ}{\mu|x|}\ge -
   \Big(\frac{2eZ}{\mu}\Big)^2
\label{hydrogen}
\ee
i.e.
$$
  T(\bA_0)-\frac{e}{\mu} W \ge 
\frac{1}{2} (\bp +\bA_0)^2 - 2\Big(\frac{eZ}{\mu}\Big)^2
 - |\bB_0| -\frac{Ce}{\mu d^2}{\bf 1}(|x|\leq 2d).
$$
We choose $\mu=4Z^2$, then using $Ce/\mu d^2 \leq e^2/4\mu$ for $\mu\leq e$
(i.e. $C\leq (\delta Z^{2/3})^2$), we get
\be
\begin{split}
 \int_\mu^\infty\cN_{0}\Big(T(\bA_0)- \frac{e}{\mu}W+\frac{e^2}{\mu}\Big)\rd e 
&\leq \int_\mu^\infty\cN_{0}\Big(\frac{1}{2} (\bp +\bA_0)^2 
 - |\bB_0| +\frac{e^2}{4\mu}\Big)\rd e \cr 
&\leq C\int_0^\mu \rd e\int_{\bR^3}
    (|\bB_0|-e^2/4\mu)_+^{3/2}\cr
&\leq C\mu^{1/2}\int_{\bR^3} \bB_0^2 .
\label{secondclr}
\end{split}
\ee

Note that if $Z\al^2\leq \kappa$ with some sufficiently small
universal constant $\kappa$, then
the magnetic energy terms in \eqref{firstclr} and
\eqref{secondclr} can be controlled by the corresponding
term in \eqref{h1}.
Combining the
 estimates \eqref{h1}, \eqref{running},
\eqref{firstclr} and \eqref{secondclr}
 we obtain
\be
\label{h1final}
  H_{N,Z}^1(\bA)\ge 
   - CZ^{5/2}d^{1/2} -Cd^{-2}
\ee
and  Lemma \ref{lm:h1} follows.
$\Box$

\section{Removing the magnetic field}\label{sec:h0}

{\it Proof of Lemma \ref{lm:removefield}.} We start with
two preparations. In Section \ref{sec:N}
we give an upper bound for the number of electrons $N$
in the truncated model described by $H_{N,Z}^0(\bA)$. In Section \ref{sec:red}
we then reduce the problem to a one-body semiclassical statement 
on boxes. The semiclassical problem will be investigated
in Section \ref{sec:sc} and this will complete the
proof of Lemma \ref{lm:removefield}.

\subsection{Upper bound on the number of electrons $N$}\label{sec:N}

Let
$$
   E^0_{N,Z}(\bA):= 
  \inf\Big\{ \langle\Psi, H_{N,Z}^0(\bA)\Psi\rangle \; :
\;  \Psi\in \bigwedge_1^N \big(C_0^\infty(\bR^3)\otimes \bC^2\big),
 \; \|\Psi\|=1\Big\}
$$
be the ground state energy of the truncated Hamiltonian 
$H_{N,Z}^0(\bA)$ defined
in \eqref{h0h1}. The following lemma shows that 
we can assume
$N\leq CZ$ when taking the infimum over $N$ in
\eqref{eq:remove}.
The proof is a slight modification of the proof of the 
Ruskai-Sigal theorem as presented in \cite{CFKS}.
We note that the original proof was given for the non-magnetic case
and it can be trivially extended to the Schr\"odinger operator
with a magnetic field but not to the Pauli operator.
This is because a key element of the proof, the standard lower
bound on the hydrogen atom, $-\Delta - Z/|x|\ge -Z^2/4$,
is valid if $-\Delta=\bp^2$ replaced by $(\bp+\bA)^2$
but there is no lower bound for the ground state energy of
the hydrogen atom with the Pauli kinetic energy that 
is independent of the magnetic field. However, for the truncated
Coulomb potential the trivial lower bound can be used.

\begin{lemma} There exist universal constants $c$ and $C$ such that
for any fixed $\bA\in \cA_0$ and $Z$ we have
$$
    E^0_{N,Z}(\bA) = E^0_{N-1,Z}(\bA)
$$
whenever $N\ge CZ$ and  $Z^{-1/6}\leq \delta \leq c$. In particular
\be
 \inf_N \inf_{\bA\in \cA_0} E_{N,Z}^0(\bA) 
 = \inf_{N\leq CZ} \inf_{\bA\in \cA_0} E_{N,Z}^0(\bA)
\label{infbound}
\ee
if  $Z^{-1/6}\leq \delta \leq c$.
\end{lemma}

{\it Proof.}   We mostly follow the proof of Theorem 3.15 in \cite{CFKS}
and we will indicate only the necessary changes.  
For any $x=(x_1, x_2, \ldots x_N)\in \bR^{3N}$ we
define 
\be
\begin{split}
x_\infty(x) & :=\max \{ |x_i|, \, : \, i=1,2, \ldots , N\}\cr
   A_0 &:=\{x\; : \; |x_j|< \varrho  \;\; \forall \;\; j=1,2,\ldots N\}\cr
  A_i& : = \{ x\; :\; |x_i|\ge (1-\zeta)x_\infty(x), \; x_\infty(x)>
\frac{\varrho}{2}\}
\nonumber
\end{split}
\ee
for some fixed positive $\varrho$ and $\zeta < 1/2$ to be chosen later. 
According to Lemma 3.16 in \cite{CFKS}, there is partition of unity 
$\{ J_i\}_{i=0, 1, \ldots N}$, with $\sum_i J_i^2\equiv1$,
$\mbox{supp}\, J_i\subset A_i$ such that the gradient estimates
\be
\begin{split}
L(x)=\sum_{i=0}^N |\nabla J_i(x)|^2 \leq \frac{CN^{1/2}}{\varrho^2} &
 \qquad \mbox{if $x\in A_0$}\cr
L(x)=\sum_{i=0}^N |\nabla J_i(x)|^2 \leq \frac{CN^{1/2}}{\varrho x_\infty(x)} &
 \qquad \mbox{if $x\in A_j$, $j\ge1$}
\nonumber
\end{split}
\ee
hold with a suitable universal constant $C$. Moreover, $J_0$ is symmetric
in all variables, while $J_i$, $i\ge 1$, is symmetric in all
variables except $x_i$.

We subtract the local field energy that is an irrelevant
constant, i.e. define 
$$
 H_N := H_{N,Z}^0(\bA) - 
 \frac{1}{16\pi\al^2}\int_{B_{3r}}|\nabla\otimes \bA|^2
$$
and $E_N = \inf\mbox{Spec}\; H_N$. We will show
that $E_N=E_{N-1}$ for $N\ge CZ$. By removing one electron to 
infinity, clearly $E_N\le E_{N-1}\leq0$; here we used the fact
that $\bA$ is compactly supported.

By the IMS localization
\be
   H_N
 = J_0(H_N-L)J_0 + \sum_{i=1}^N J_i(H_N-L)J_i.
\label{ims1}
\ee
In the first term we use that on the support of $\theta_0$ we
  have $-Z|x|^{-1}\geq -Zd^{-1}$. Hence
\be
   J_0(H_N-L)J_0 \ge J_0\Big( - CZNd^{-1} - CNd^{-2} + \frac{N(N-1)}{4\varrho}
  -\frac{CN^{1/2}}{\varrho^2}\Big)J_0.
\label{fi}
\ee
Choosing $\varrho =8d$
we see that $J_0(H_N-L)J_0\ge 0$ if $N\ge CZ$ with a constant $C$ 
if $\delta\ge CZ^{-2/3}$.

To estimate the terms $J_i(H_N-L)J_i$ for $i\neq 0$, we define
\be
\begin{split}
   H_{N-1}^{(i)}: = &  \sum_{j=1\atop j\neq i}^N 
\Bigg[\theta_0
 \Big (T(\bA) -
 \frac{Z}{|x|} - \big(|\nabla\theta_0|^2
 +|\nabla\theta_1|^2\big)\Big)\theta_0\Bigg]_j + \sum_{k<j\atop k,j \neq i}
 \frac{1}{|x_k-x_j|}.\nonumber
\end{split}
\ee
On the support of $J_i$ we have $|x_i|\ge \varrho/4=2d$, so
$\nabla\theta_0$ and $\nabla\theta_1$ vanish.
Then we can  estimate
\be
\begin{split}
J_i(H_N-L)J_i \ge & J_i\Bigg(H_{N-1}^{(i)} - \frac{Z}{|x_i|}
+ \frac{N-1}{2x_\infty(x)} - \frac{CN^{1/2}}{x_\infty(x)\varrho}\Bigg)J_i\cr
\ge & J_i\Bigg( E_{N-1} + \frac{1}{|x_i|}\Big[ \frac{N-1}{2}(1-\zeta)
  - Z -  \frac{CN^{1/2}Z^{1/3}}{\delta}\Big]\Bigg)J_i\cr
\ge & J_iE_{N-1}J_i\label{sec}
\end{split}
\ee
if $N\ge CZ$ and $N$ is large. Thus we conclude from \eqref{ims1},
\eqref{fi} and \eqref{sec} that $E_N\ge E_{N-1}$ if $N\ge CZ$.
$\Box$

\subsection{Reduction to a one-body problem}\label{sec:red}

We start by presenting an abstract lemma whose proof is given in Appendix
\ref{sec:abstr}.

\begin{lemma}\label{lm:abstr} 
Let $\fh$ be a one-particle operator on $\cH=L^2(\bR^3)$ and let $W$
be a two-particle  operator defined on $\cH\wedge \cH$.
 We assume that the domains
of $\fh$ and $W$ include the $C_0^\infty$ functions.
Let $\theta\in C^\infty(\bR^3)$ with compact support
$\Omega: =\mbox{supp}\, \theta$. Then
\be
\begin{split}\label{eq:abstr}
   \inf\Bigg\{ \Big\langle \Psi, &
\Big[\sum_{i=1}^N  \theta_i\fh_i\theta_i +  \sum_{1\leq i < j\leq N}
 \theta_i\theta_j W_{ij}\theta_j\theta_i\Big]
\Psi \Big\rangle\; :\; \Psi\in \bigwedge_1^N C_0^\infty(\bR^{N}),
\; \|\Psi\|=1\Bigg\}\cr
& \ge
   \inf_{n\leq N} \inf
 \Bigg\{ \Big\langle \Phi, 
\Big(\sum_{i=1}^n  \fh_i + \sum_{1\leq i< j\leq n} W_{ij}\Big)
\Phi \Big\rangle\; : \; \Phi\in\bigwedge_1^n
  C_0^\infty(\Omega), \; \|\Phi\|=1\Bigg\}\, ,
\end{split}
\ee
where $\fh_i$ denotes the operator  $\fh$ acting
on the $i$-th component of  the tensor product, and similar
convention is used for the two-particle operators.
The same result holds with obvious changes if 
$\cH= L^2(\bR^3)\otimes \bC^2$. 
\end{lemma}

\bigskip

To continue the proof of Lemma \ref{lm:removefield},  we
first localize $H^0_{N,Z}(\bA)$ onto a ball $B_r$ of radius $r=DZ^{-1/3}$
(see \eqref{rchoice}) and we also localize the magnetic field
as  in Section \ref{sec:tooth}. 
We introduce smooth cutoff functions $\chi_0$ and $\chi_1$ with
$$
  \chi_0^2 + \chi_1^2\equiv 1, \;\; \mbox{supp}\; \chi_0 \subset B_{2r},
\quad \chi_0\equiv 1 \;\;\mbox{on $B_r$}, 
\quad |\nabla \chi_0|, \;|\nabla \chi_1|\leq Cr^{-1}\, .
$$
We get
\be
\begin{split}
  H^0_{N,Z}(\bA) \ge &   \sum_{i=1}^N \theta_0\chi_0 e^{-i\varphi_0}
  \Big[ T_i(\bA_0) - \frac{Z}{|x_i|}\Big] e^{i\varphi_0}
\chi_0\theta_0  + 
\sum_{i<j}  \frac{1}{|x_i-x_j|} \cr
& + \frac{1}{C\al^2}\int_{\bR^3} \bB^2_0
  - CNd^{-2} - CNZr^{-1}
\end{split}
\ee
using that the new
localization error $|\nabla\chi_1|^2+|\nabla\chi_0|^2
 \leq Cr^{-2}\leq Cd^{-2}$ and 
that $-Z/|x|\ge -Zr^{-1}$ on the support of $\chi_1$.
We also used  \eqref{bb0}.

Let $A_{d,r}=\{ x\; : \; d\leq |x|\leq r\}\subset \bR^3$.
Using \eqref{infbound},  the positivity of the Coulomb repulsion
$|x_i-x_j|^{-1}>0$ and
Lemma \ref{lm:abstr} with $\theta:=\theta_0\chi_0$
 we obtain
\be
\begin{split}\label{psi}
   \inf_N \inf_{\bA\in\cA_0} H_{N,Z}^0(\bA) & \ge  \inf_{N\leq CZ} 
\inf_{\bA_0\in \cA_0}
  \Bigg\{ \inf_{\Psi} 
  \Big\langle \Psi, \Big[\sum_{i=1}^N  \Big[T(\bA_0)-
\frac{Z}{|x|}\Big]_i +\sum_{i<j}\frac{1}{|x_i-x_j|}\Big]_{A_{d,r}}\Psi
\Big\rangle\cr
& \;\;\; +\frac{1}{C\al^2}\int_{\bR^3} \bB_0^2\Bigg\}
  - CZ^{5/3}\delta^{-2} - CZ^{7/3} D^{-1},
\end{split}
\ee
where the infimum is over all antisymmetric wave functions $\Psi\in
\bigwedge_1^N C_0^\infty(A_{d,r})\otimes \bC^2$ with $\|\Psi\|_2=1$.
 The notation
$[H]_Q$ indicates the $N$-particle operator $H$ 
with Dirichlet boundary condition on the domain 
$Q^N\subset \bR^{3N}$.

\bigskip

We  define 
$$
   D(f,g):=\frac{1}{2}\iint_{\bR^3\times\bR^3}
\frac{f(x)g(y)}{|x-y|}\rd x \rd y.
$$
\begin{lemma}\label{lm:lo} There is a universal constant $C_0>0$ such
that for any $\Psi\in
\bigwedge_1^N C_0^\infty(\bR^3)\otimes \bC^2$ with $\|\Psi\|_2=1$, for 
any nonnegative function $\varrho: \bR^3\to \bR$ with
$D(\varrho, \varrho)<\infty$, for any $\bA\in \cA_0$ and for any $\e>0$ we have
\be
\begin{split}
   \Big\langle \Psi, \Big[\e\sum_{i=1}^N  T_i(\bA) +
\sum_{i<j}\frac{1}{|x_i-x_j|}\Big]\Psi
\Big\rangle & + C_0\int_{\bR^3} \bB^2 \cr 
&\ge
- D(\varrho, \varrho)
  +  \Big\langle \Psi,\sum_{i=1}^N \big(
\varrho\ast |x_i|^{-1}\big)\Psi\Big\rangle 
-C\e^{-1}N.
\label{lo}
\end{split}
\ee
\end{lemma}

{\it Proof.}
By the Lieb-Oxford inequality \cite{LO} and by
the positivity of the quadratic form $D(\cdot, \cdot)$
\be
\begin{split}
  \Big\langle \Psi,\sum_{i<j}\frac{1}{|x_i-x_j|}\Psi\Big\rangle &
 \ge  D(\varrho_\Psi, \varrho_\Psi)
  - C\int_{\bR^3} \varrho_\Psi^{4/3} \cr  
& \ge  - D(\varrho, \varrho)
  +  \Big\langle \Psi,\sum_{i=1}^N (\varrho\ast |x_i|^{-1})\Psi\Big\rangle 
- C\int_{\bR^3} \varrho_\Psi^{4/3}
\end{split}
\ee
where $\varrho_\Psi(x)$ is the one-particle density of $\Psi$.

The error term is controlled by the following kinetic
energy inequality for the Pauli operator
\be
   \Big\langle \Psi, \Big[\sum_{i=1}^N  T(\bA)_i \Big]\Psi\Big\rangle
  \ge c\int_{\bR^3} \min\{ \varrho_\Psi^{5/3}, \gamma \varrho_\Psi^{4/3}\}
   -  \gamma\int_{\bR^3} \bB^2
\label{kinpauli}
\ee
with some positive universal constant $c$ and for any $\gamma>0$.
For the proof of \eqref{kinpauli} use the magnetic Lieb-Thirring
inequality
$$
   \Big\langle \Psi, \Big[\sum_{i=1}^N  [T(\bA)-U]_i \Big]\Psi\Big\rangle
\ge -C\int_{\bR^3} U^{5/2} - C\gamma^{-3}\int_{\bR^3} U^4 - 
\gamma\int_{\bR^3} \bB^2.
$$
With the choice $U = \beta\min \{ \varrho_\Psi^{2/3}, \gamma
\varrho_\Psi^{1/3}\}$ we can ensure that
$\frac{1}{2} U\varrho_\Psi \ge CU^{5/2} + C\gamma^{-3} U^4$
if $\beta$
is sufficiently small  (independent of $\gamma$)
and this proves \eqref{kinpauli}.

Thus
\be
\begin{split}
\int_{\bR^3} \varrho_\Psi^{4/3}& \leq \gamma^{-1} \int_{\bR^3} 
 \min\{ \varrho_\Psi^{5/3}, \gamma \varrho_\Psi^{4/3}\} +
\gamma \int_{\bR^3} \varrho_\Psi \cr
& \leq (c\gamma)^{-1} 
\Big\langle \Psi, \Big[\sum_{i=1}^N  T(\bA)_i \Big]\Psi\Big\rangle
 +c^{-1}\int_{\bR^3} \bB^2 + \gamma N
\end{split}
\ee
so choosing $\gamma = C\e^{-1}$ with a sufficiently large
constant $C$, we obtain \eqref{lo}. $\Box$.

\bigskip

Using Lemma \ref{lm:lo} we can continue the estimate \eqref{psi}
(with writing $\bA$ instead of $\bA_0$ in the infimum) as 
\be
\begin{split}\label{psi2}
   \inf_N \inf_\bA H_{N,Z}^0(\bA) 
&  \ge  (1-\e)\inf_{N\leq CZ} \inf_{\bA\in \cA_0}
  \Bigg\{\inf_{\Psi} 
  \Big\langle \Psi, \Big[\sum_{i=1}^N  [T(\bA) + W]_i\Big]_{A_{d,r}}\Psi
\Big\rangle +\frac{1}{C\al^2}\int_{\bR^3} \bB^2\Bigg\}\cr
&\;\;\;\; 
-  D(\varrho, \varrho)
 -C\e^{-1} Z - CZ^{5/3}\delta^{-2}- CZ^{7/3} D^{-1}
\end{split}
\ee
with
$$
 W(x): = \frac{1}{1-\e}\Big[- \frac{Z}{|x|} +\varrho\ast |x|^{-1}\Big]
$$
and assuming that $\al\leq \al_0$ with some small universal $\al_0$.
\bigskip

We now perform a  rescaling: $x= Z^{-1/3}X$,
$\bp = Z^{1/3}\bP$
and 
$$
   \wt\bA(X)= Z^{-2/3}\bA(Z^{-1/3}X),
\qquad  \wt\bB(X) =\nabla \times \wt\bA(X)= Z^{-1} \bB(Z^{-1/3}X)
$$
Introducing $h=Z^{-1/3}$ and $T_h(\wt\bA): = [(h\bP+\wt\bA)\cdot\bsigma]^2$,
we obtain that 
the kinetic energy changes as
$$
    [(\bp+\bA)\cdot \bsigma]^2 
    = Z^{4/3} [(Z^{-1/3}\bP+\wt\bA)\cdot\bsigma]^2 = Z^{4/3}T_h(\wt \bA)
$$
and
the field energy changes as
$$
    \int_{\bR^3} \bB^2(x) \rd x = Z\int_{\bR^3} \wt\bB^2(X) \rd X.
$$
The new potential energy is
$$
  \wt W(X) = Z^{-4/3}W(Z^{-1/3}X)=
\frac{1}{1-\e}\Bigg[-\frac{1}{|X|} + \wt \varrho
\ast |X|^{-1}\Bigg],
$$
where $\wt\varrho(X) = Z^{-2}\varrho (Z^{-1/3}X)$ and $D(\wt\varrho,
\wt \varrho)= Z^{-7/3} D(\varrho, \varrho)$.
After rescaling,  we get from \eqref{psi2}
\be
\begin{split}\label{psi3}
   \inf_N & \inf_{\bA\in\cA_0} H_{N,Z}^0(\bA)  \cr
&  \ge  (1-\e) Z^{4/3} \inf_{N\leq CZ} \inf_{\wt \bA\in \cA_0}
  \Bigg\{\inf_{\Psi} 
  \Big\langle \Psi, \Big[\sum_{i=1}^N  [T_h(\wt\bA) + \wt W]_i
\Big]_{ A_{\delta, D}}\Psi
\Big\rangle +\frac{h^{-2}}{CZ\al^2}\int_{\bR^3} \wt \bB^2\Bigg\}\cr
&\;\;\;\; 
- Z^{7/3} D(\wt\varrho, \wt\varrho)
 -C\e^{-1} Z - CZ^{5/3}\delta^{-2}-CZ^{7/3} D^{-1},
\end{split}
\ee
where $ A_{\delta,D} =\{ X\; : \; \delta\leq |X|\leq D\}$
and $\inf_\Psi$ denotes infimum over all normalized antisymmetric functions.
Using \eqref{sc3} from Theorem
 \ref{lm:sc} and the fact that $Z\al^2\leq\kappa$,
 we get
\be
\begin{split}\label{psi4}
   \inf_N \inf_\bA H_{N,Z}^0(\bA) 
&  \ge  (1-\e) Z^{4/3}  \Tr \big[ 
(-h^2\Delta +\wt W)_{A_{\delta,D}}]_- - CZ^{13/6}D^3\delta^{-13/4}\cr
&\;\;\;\; 
- Z^{7/3} D(\wt\varrho, \wt\varrho)
 -C\e^{-1} Z - CZ^{5/3}\delta^{-2}-CZ^{7/3} D^{-1}
\end{split}
\ee 
assuming $\delta \ge Z^{-2/9}$.
By standard semiclassical result for Coulomb-like potentials
(see e.g. the result in Section V.2 of \cite{L}):
\be
   \Tr \big[ (-h^2\Delta +\wt W)_{A_{\delta,D}}]_-
\ge 
 \Tr \big[ -h^2\Delta +\wt W]_- \ge -C_{sc}h^{-3}\int_{\bR^3} \wt W_-^{5/2}
  - Ch^{-3+1/10}.
\label{stsc}
\ee
where $C_{sc}=2/(15\pi^2)$ is the Weyl constant in semiclassics.
The $\frac{1}{10}$ exponent in the error term is far from
being optimal; the methods developed to
prove the Scott correction can yield an exponent up to
one (see Remark~1 after Theorem \ref{thm:main1}).

Taking the optimal $\wt\varrho$ to be the Thomas-Fermi
density for $Z=1$ $\wt\varrho = \varrho_{{\rm TF}}$
 (see, e.g. Section II of \cite{L}) and
defining the Thomas-Fermi constant as
$$
   c_{{\rm TF}} := D(\varrho_{{\rm TF}}, \varrho_{{\rm TF}}) +
  C_{SC}\int_{\bR^3}\Big[-\frac{1}{|X|} + \varrho_{{\rm TF}}
\ast |X|^{-1}\Big]_-^{5/2} 
$$
we get
\be
\begin{split}\label{psi5}
   \inf_N \inf_\bA H_{N,Z}^0(\bA) 
&  \ge  (1-\e)^{-3/2} Z^{7/3} \Bigg( - D(\wt\varrho, \wt\varrho)-
C_{sc}\int_{\bR^3}\Big[-\frac{1}{|X|} + \wt \varrho
\ast |X|^{-1}\Big]_-^{5/2} \Bigg) \cr
&\;\;\;\;- CZ^{7/3 -1/30} - CZ^{13/6}D^3\delta^{-13/4}
 -C\e^{-1} Z - CZ^{5/3}\delta^{-2}-CZ^{7/3} D^{-1} \cr
&\ge -(1-\e)^{-3/2}c_{{\rm TF}} Z^{7/3} -CZ^{7/3 -1/30}
 -C\e^{-1} Z - CZ^{5/3}\delta^{-2} - CZ^{7/3}D^{-1}\cr
&\ge   -c_{{\rm TF}} Z^{7/3} -C Z^{7/3}\big[
 Z^{-1/30} + D^{-1}\big]\cr
\end{split}
\ee
where we optimized for $\e$  and we used that $D\leq Z^{1/24}\delta^{13/16}$
and $ Z^{-1/6}\leq\delta\leq 1$.
This completes the proof of Lemma \ref{lm:removefield}. $\Box$

\bigskip

\section{Semiclassics: proof of Theorem \ref{lm:sc}}\label{sec:sc}

We present the Schr\"odinger and Pauli cases in parallel.
We prove the statement with Dirichlet boundary conditions
\eqref{sc3} in details and in Section \ref{sec:reduce}
we comment on the necessary changes
for the proof of the \eqref{sc1}.  
The potential $V$ is defined only on $\Omega$, but we extend
it to be zero on $\bR^3\setminus \Omega$
we continue to denote by $V$ its extension.

\subsection{Localization onto boxes}\label{sec:loc}

We choose a length $L$ with
$h\le L\le \frac{1}{3}h^{1/2}$.
Let $\Omega_L= \Omega+B_L$ be the $L$-neighborhood of $\Omega$.
Let $Q_k=\{ y\in \bR^3\; : \; \| y-k\|_\infty < L/2\}$
with $k\in (L\bZ)^3\cap \Omega_L$ denote  a non-overlapping
covering  of $\Omega$
 with boxes of size $L$.
In this section
the index $k$ will always run over the set $(L\bZ)^3\cap \Omega_L$.
Let $\xi_k$ be a partition of unity, $\sum_k \xi^2_k\equiv 1$, subordinated to 
the collection of boxes $Q_k$, such that
$$
    \mbox{supp}\;\xi_k \subset (2Q)_k, \qquad
    |\nabla \xi_k|\leq CL^{-1}
$$
where $(2Q)_k$ denotes the cube of side-length $2L$
with center $k$. Let $\wt\xi_k$  be a cutoff
function such that $\wt\xi_k \equiv 1$ on $(2Q)_k$ (i.e. on the
support of $\xi_k$),
$\mbox{supp}\, \wt\xi_k\subset \wt Q_k:=(3Q)_k$
and $|\nabla \wt\xi_k|\leq CL^{-1}$.

Let $\langle\bA\rangle_k = |\wt Q_k|^{-1}\int_{\wt Q_k} \bA$.
  Similarly to \eqref{a0}, we define
$\bA_k: = (\bA- \langle\bA\rangle_k)\wt\xi_k$ and
$\bB_k : = \nabla\times\bA_k$,
then
\be
\int_{\bR^3} \bB_k^2 \leq C\int_{\wt Q_k} |\nabla \otimes\bA|^2
\label{ba}
\ee
as in \eqref{bb0}.
{F}rom the IMS localization with $\psi_k$ satisfying 
$h\nabla\psi_k = \langle \bA_k\rangle$ we have
\be
\begin{split}
\Tr\Big[  [ T_h(\bA) -V]_\Omega \Big]_- + h^{-2}\int_{\bR^3} \bB^2 
= & \inf_\gamma^*  \Tr\Big( \gamma [T_h(\bA)-V]\Big)
 +h^{-2}\int_{\bR^3}|\nabla \otimes\bA|^2 \cr
 \ge &  \inf_\gamma^*\sum_{k\in(L\bZ)^3\cap \Omega_L} \cE_k(\gamma)
\end{split}\label{qk}
\ee
with
$$
  \cE_k(\gamma):= \Tr \Big[ \gamma \xi_k e^{-i\psi_k}
[T_h(\bA-\langle\bA\rangle_k)-V]e^{i\psi_k}\xi_k -
\gamma|h\nabla\xi_k|^2\Big]
+c_0h^{-2}\int_{\wt Q_k}|\nabla \otimes\bA|^2
$$
with some universal constant $c_0$.
Here $\inf_\gamma^*$ denotes infimum 
 over all density matrices $0\leq \gamma\leq 1$ that are
supported on $\Omega$, i.e. they are operators on $L^2(\Omega)\otimes \bC^2$.
We also used $\int_{\bR^3}\bB^2= \int_{\bR^3}|\nabla \otimes\bA|^2$
and we  reallocated the second integral. We introduce the notation
$$
  \cF_k:=c_0h^{-2}\int_{\wt Q_k}|\nabla \otimes\bA|^2.
$$

\subsection{Apriori bound on the local field energy}

In case of the Pauli operator,
for each fixed   box $\wt Q_k$ we apply  the magnetic Lieb-Thirring inequality
\cite{LLS} together with \eqref{ba} and
with the bound $\| V \|_\infty\leq K$ to obtain that for any density matrix
$\gamma$
\be
\begin{split}\label{scr}
\cE_k(\gamma) &\ge  \Tr\Big[  [T_h(\bA_k)-V- Ch^2L^{-2}]_{\wt Q_k} \Big]_- 
+\cF_k\cr
 & \ge - Ch^{-3}\int_{\wt Q_k}[ V+ Ch^2L^{-2}]^{5/2}
  - C \Big( \int_{\wt Q_k} [ V+ Ch^2L^{-2}]^4\Big)^{1/4}
    \Big( h^{-2}\int_{\wt Q_k} \bB_k^2\Big)^{3/4}+\cF_k\cr
& \ge - C\Big[h^{-3}K^{5/2}L^3 + h^2L^{-2} + 
 K^4L^3 +  h^8L^{-5}\Big]
 -  \frac{c_0}{2}\, h^{-2}\int_{\wt Q_k} |\nabla\otimes \bA|^2+\cF_k\cr
&\ge - Ch^{-3}K^{5/2}L^3+\frac{1}{2}\cF_k
\end{split}
\ee
using $h\le L$ and $1\leq K\le Ch^{-2}$.
In the Schr\"odinger case we use the usual Lieb-Thirring
inequality \cite{LT} that holds with a magnetic field as well. 
The estimate \eqref{scr} is then valid even without the third term
in the second line.

Let $S\subset (L\bZ)^3\cap  \Omega_L$ 
denote the set of those $k$ indices 
such that 
\be
\begin{split}
 \cF_k
& \leq  Ch^{-3}K^{5/2}L^3.
\end{split}\label{aprio}
\ee
holds with some large constant $C$. In 
 particular  
\be
  \cE_k(\gamma)\ge 0, \quad \mbox{for all $k\not\in S$ and for
any $\gamma$}.
\label{cebig}
\ee

\subsection{Improved bound}

We use the Schwarz inequality
in the form
$$
T_h(\bA-\langle\bA\rangle_k) \ge -(1-\e_k)h^2\Delta - C\e_k^{-1}(\bA-\langle
\bA\rangle_k)^2,
$$
with some $0<\e_k<\frac{1}{3}$. 
We  have 
for any $\gamma$ supported on $\Omega$ that
\be
\begin{split}
 \cE_k(\gamma)\ge &
\Tr\Big[{\bf 1}_\Omega\xi_k 
[-(1-2\e_k)h^2\Delta-V-Ch^2L^{-2}]\xi_k{\bf 1}_\Omega\Big]_-\cr & +
\Tr \Big[{\bf 1}_{\wt Q_k} [-\e_k h^2\Delta-C\e_k^{-1}(\bA-\langle
\bA\rangle_k)^2] {\bf 1}_{\wt Q_k}\Big]_- +\cF_k\cr
\end{split}
\ee
We will show at the end of the section that
\be
\begin{split}
  \Tr\Big[ {\bf 1}_\Omega\xi_k [-(1&-2\e_k)h^2\Delta -V-Ch^2L^{-2}]\xi_k
{\bf 1}_\Omega\Big]_-\cr
  &\ge \Tr \Big[ {\bf 1}_\Omega\xi_k(-h^2\Delta -V)\xi_k{\bf 1}_\Omega\Big]_- 
- Ch^{-3}K^{5/2}
  (\e_k + h^2L^{-2}) |\wt Q_k|.
\label{stand}
\end{split}
\ee
Using \eqref{cebig} and \eqref{stand},
\be
\begin{split}\label{schw}
\inf^*_\gamma\sum_k \cE_k(\gamma)\ge & \inf^*_\gamma\sum_{k\in S} 
\cE_k(\gamma) \cr
\ge&
 \sum_k \Tr\Big[ {\bf 1}_\Omega\xi_k [-h^2\Delta-V]\xi_k{\bf 1}_\Omega\Big]_-
 + \sum_{k\in S} \cD_k\cr
\ge&  
  \sum_k \inf_{\gamma_k}\Tr\Big[ \xi_k \gamma_k\xi_k{\bf 1}_\Omega
[-h^2\Delta-V]{\bf 1}_\Omega\Big]
+ \sum_{k\in S} \cD_k\cr
\ge & \; \Tr \big[ (-h^2\Delta-V)_{\Omega}
\big]_- +\sum_{k\in S} \cD_k
\end{split}
\ee
with
\be
 \cD_k:=
\Tr \Big[  [-\e_k h^2\Delta-
C\e_k^{-1}(\bA-\langle\bA\rangle_k)^2]_{\wt Q_k}\Big]_- 
- Ch^{-3}K^{5/2}|\wt Q_k| (\e_k + h^2L^{-2})
+\cF_k .
\label{ce}
\ee
In the last step in \eqref{schw} 
we used that for any collection of density matrices
$\gamma_k$, the density matrix $\sum_k {\bf 1}_\Omega\xi_k \gamma_k
\xi_k{\bf 1}_\Omega$ is
 admissible  in the variational
principle 
\be
\begin{split}
\Tr \big[ ( & -h^2\Delta  -V)_\Omega
\big]_- =
\inf\Big\{\Tr\gamma\big[-h^2\Delta-V\big]\; :
 \; 0\leq \gamma\leq 1, \; \mbox{supp}\,\gamma\subset \Omega
\Big\}.
\end{split}\label{varprin}
\ee

We estimate $\cD_k$ for $k\in S$ as follows
\be
\begin{split}
\cD_k
\ge & -C\e_k^{-4} h^{-3}
 \int_{\wt Q_k} (\bA-\langle\bA\rangle_k)^5 
- Ch^{-3}K^{5/2}|\wt Q_k|  (\e_k + h^2L^{-2})
 +\cF_k\cr
\ge &\;  \cF_k -C\e_k^{-4}h^{2}L^{1/2}\cF_k^{5/2}
 - Ch^{-3}K^{5/2} |\wt Q_k| (\e_k + h^2L^{-2}).
\end{split}
\ee
In the first step we used Lieb-Thirring inequality, in the second
step  H\"older and Sobolev inequalities
in the form
$$
   \int_{\wt Q_k} (\bA-\langle\bA\rangle_k)^5 \leq
  CL^{1/2} 
\Big(\int_{\wt Q_k}|\nabla\otimes \bA|^2\Big)^{5/2}.
$$
We choose
$$
  \e_k = hL^{-1/2}K^{-1/2}\cF^{1/2}_k
$$
and using the apriori bound \eqref{aprio}, we see that
$$
  \e_k \leq 
    Ch^{-1/2}L K^{3/4}.
$$
Thus, assuming
\be
   L \leq c h^{1/2}K^{-3/4}
\label{Lcond}
\ee
with a sufficiently small constant $c$,
we get $\e_k\leq 1/3$.
With this choice of $\e_k$, and recalling $|\wt Q_k| = 9|Q_k|=9L^3$, we have
\be
\begin{split}
  \cD_k & 
\ge \cF_k - C h^{-2}L^{5/2}K^2\cF_k^{1/2} - Ch^{-3}K^{5/2}L^3 h^2L^{-2}\cr
&\ge - C h^{-3} L^3 K^{5/2}\Big( h^{-1}L^2 K^{3/2} + h^2L^{-2}\Big).
\end{split}\label{d1}
\ee
If we choose $L= h^{3/4} K^{-3/8}$, then
$$
 \cD_k\ge - C h^{-3} L^3 K^{5/2} \Big(h K^{3/2}\Big)^{1/2}.
$$
 This choice is allowed by \eqref{Lcond}
if $K\leq ch^{-2/3}$. If $ch^{-2/3}\leq K \leq h^{-2}$, then we choose
$L = ch^{1/2}K^{-3/4}$ and we get from \eqref{d1}
$$  
\cD_k \ge - C h^{-3} L^3 K^{5/2} (1+ hK^{3/2}).
$$
Combining these two inequalities, we get that
\be
\cD_k \ge - C h^{-3} L^3 K^{5/2}\big(h K^{3/2}\big)^{1/2} 
\big[1+\big(h K^{3/2}\big)^{1/2}  \big]
\label{dest}
\ee
always holds.
Summing up \eqref{dest} for all $k$ and using that 
$$
  \sum_{k\in (L\bZ)^3\cap \Omega_L} L^3 \leq C |\Omega_{3L}|\leq 
C|\Omega_{\sqrt{h}}|
$$
(recall that $\Omega_{\sqrt{h}}$ is a $\sqrt{h}$-neighborhood of $\Omega$ and
$3L\leq h^{1/2}$),
we obtain
from  \eqref{schw} and \eqref{dest} 
\be
\begin{split}\label{schw3}
\inf^*_\gamma\sum_k \cE_k(\gamma)\ge & 
\Tr \Big[(-h^2\Delta-V)_\Omega\Big]_- - Ch^{-3}K^{5/2}|\Omega_{\sqrt{h}}|
\big(h K^{3/2}\big)^{1/2} 
\big[1+\big(h K^{3/2}\big)^{1/2}  \big]
\bigskip\bigskip
\end{split}
\ee
and this proves  \eqref{sc3}.

\bigskip

Finally, we prove \eqref{stand}.
 Let $\gamma$ be a trial density matrix for
the left hand side of \eqref{stand}. We can assume
that
$$
  0\ge \Tr\Big[\gamma {\bf 1}_\Omega
\xi_k [-(1-2\e_k)h^2\Delta -V-Ch^2L^{-2}]\xi_k {\bf 1}_\Omega\Big]
$$
Then 
\be
\begin{split}
 0 \ge & \Tr\Big[\gamma {\bf 1}_\Omega
\xi_k [-\frac{1}{6}h^2\Delta +K]\xi_k {\bf 1}_\Omega\Big] 
+ \Tr\Big[\gamma {\bf 1}_\Omega
\xi_k [-\frac{1}{6}h^2\Delta -V-Ch^2L^{-2}-K]\xi_k {\bf 1}_\Omega\Big]\cr
\ge  &\Tr\Big[\gamma {\bf 1}_\Omega
\xi_k [-\frac{1}{6}h^2\Delta +K]\xi_k {\bf 1}_\Omega\Big] 
 - Ch^{-3}\int_{\wt Q_k} [ V+ K + Ch^2L^{-2}]^{5/2} ,
\end{split}
\ee
where we used Lieb-Thirring inequality. Thus, using $|V|\leq K$, $h\le L$
and $K\ge 1$, we have
$$
  \Tr\Big[\gamma {\bf 1}_\Omega
\xi_k [-\frac{1}{6}h^2\Delta +K]\xi_k {\bf 1}_\Omega\Big] 
 \leq Ch^{-3}K^{5/2}|\wt Q_k|.
$$
Therefore
\be
\begin{split}
  \Tr\Big[ \gamma
{\bf 1}_\Omega\xi_k [-(1&-2\e_k)h^2\Delta -V-Ch^2L^{-2}]\xi_k
{\bf 1}_\Omega\Big]\cr
  &\ge \Tr \Big[ \gamma
{\bf 1}_\Omega\xi_k(-h^2\Delta -V)\xi_k{\bf 1}_\Omega\Big]
 - Ch^{-3}K^{5/2}
  (\e_k + h^2L^{-2}) |\wt Q_k|.
\label{stand3}
\end{split}
\ee
Now  \eqref{stand} follows
by variational principle.
$\Box$

\subsection{Reduction of  \eqref{sc1} to \eqref{sc3}}\label{sec:reduce}

We approximate $V\in L^{5/2}\cap L^4$ by a bounded potential
$\wt V$, $\| \wt V\|_\infty\leq K$,
 that is supported on a ball $B_{R/2}$
 and $\wt V \leq V$. By choosing $K$ and $R$
sufficiently large, we can make $\| V-\wt V\|_{5/2} + \| V- \wt V\|_4$
arbitrarily small. We choose a cutoff function $\chi_R$ 
that is supported on $B_{R}$, $\chi_R\equiv 1$ on $B_{R/2}$
and $|\nabla\chi_R|\leq CR^{-1}$ and let $\wt\chi_R$ satisfy
$\chi_R^2 +\wt\chi_R^2\equiv 1$.

Borrowing a small part of the kinetic energy,
by IMS localization we have 
\be
\begin{split}
   T_h(\bA) - V\ge &
    (1-\e) \chi_R [T_h(\bA) - \wt V]\chi_R \cr 
& \;\; +\e T_h(\bA) - (V- (1-\e)\wt V) - |\nabla\chi_R|^2
- |\nabla\wt\chi_R|^2
\end{split}
\ee
Using the magnetic Lieb-Thirring inequality \cite{LLS}
to estimate the second term, we get
\be
\begin{split}
   \Tr [T_h(\bA) - V]_- \ge &
    (1-\e) \Tr [( T_h(\bA) - \wt V)_{B_R}]_-\cr
& \; -C\e^{-3/2}h^{-3}\int_{\bR^3} |U|^{5/2} - 
 C\int_{\bR^3} |U|^{4}-
  \frac12h^{-2}\int_{\bR^3} \bB^2
\end{split}\label{yt}
\ee
with
$$ 
  U: =  (V-(1-\e)\wt V) + |\nabla\chi_R|^2
+ |\nabla\wt\chi_R|^2 \; .
$$
For the first term in \eqref{yt} we use \eqref{sc3} (and that
it holds even with a $1/2$ in front of $\int \bB^2$) 
and the fact that 
$$
   \Tr \big[ (-h^2\Delta - \wt V)_{B_R}\big]_- 
  \ge   
   \Tr \big[ -h^2\Delta -  V\big]_-
$$
by monotonicity, $\wt V \leq V$.
The second and the third terms  in \eqref{yt} can be made arbitrarily 
small compared with $h^{-3}$ for any fixed $\e$ 
if $R$ and $K$ are sufficiently large
and $h$ is small. Finally, choosing  
$\e$ sufficiently small, we proved  \eqref{sc1}. $\Box$

\section{Proof of the quantized field case}\label{sec:q}

For the proof of the lower bound in \eqref{mainqed}, we follow
the argument of \cite{BFG} to reduce the problem to
the classical bound \eqref{eq:main}.
We set
$$
   H_g = \al^{-1}\int_{\bR^3} |g(k)|^2|k| \sum_{\lambda= \pm}
  a_\lambda(k)^*a_\lambda(k) \rd k.
$$
to be the cutoff field energy,
then $H_f\ge H_g$ and only the modes appearing in $H_g$
interact with the electron.
 By Lemma 3 of \cite{BFG}, for any real function $f\in L^1(\bR^3)\cap
L^\infty(\bR^3)$ we have
$$
   \frac{1}{8\pi}\int_{\bR^3} f(x) |\nabla \otimes \bA(x)|^2 \rd x
\leq \al^2 \|f\|_\infty H_g
  + C\al \Lambda^4 \|f\|_1.
$$
Applying it with $f$ being the characteristic function of the ball $B_{3r}$
with $r=DZ^{-1/3}$ (the radius of the ball is here chosen
  differently from \cite{BFG}) and  using $Z\al^2\leq \kappa$ we get
$$
   H_f\ge H_g \ge \Big(\frac{Z\al^2}{\kappa}\Big) H_g
   \ge \frac{Z}{8\pi \kappa}\int_{B_{3r}} |\nabla \otimes \bA(x)|^2\rd x
   - C\kappa^{-1}\al \Lambda^4 D^3.
$$
Setting $\wt\al= (\kappa/Z)^{1/2}$, i.e. $Z\wt \al^2=\kappa$,
we have for $\kappa$ sufficiently small
$$
   E_{N,Z}^{\rm qed}\ge E_{N,Z,\wt \al}
 - C\kappa^{-1}\al \Lambda^4 D^3,
$$
where $E_{N,Z,\wt \al}$ is the ground state
energy of the Hamiltonian \eqref{ham} with fine structure
constant $\wt \al$. Applying \eqref{eq:main} to this Hamiltonian,
 we
get
$$
   E_{Z}^{\rm qed}\ge E_Z^{\rm nf} - CZ^{\frac{7}{3}}D^{-1}
 - C\kappa^{-1}\al \Lambda^4 D^3
$$
whenever $1\leq D\leq Z^{\frac{1}{63}}$. Writing $D= Z^{\gamma}$ and
applying the upper bound \eqref{Lam} on $\Lambda$, we obtain
the lower bound in \eqref{mainqed}. $\Box$

\appendix

\section{Proof of Lemma \ref{lm:abstr}}\label{sec:abstr}

Let the function $\chi(x)\in C^\infty(\bR^3)$ 
be defined such that $\theta^2(x)+\chi^2(x)
\equiv 1$.
For any subset $\al\subset\{1, 2, \ldots, N\}$ we denote
by $x_\al$ the collection of variables $\{ x_i \; : \; i\in \al\}$
and define
$$
  \Theta_\al=\Theta_\al(x_\al):= \prod_{i\in \al}\theta(x_i),
\quad 
 \Xi_\al=\Xi_\al(x_\al):= \prod_{i\in \al}\chi(x_i).
$$
We set the notation $\al^c=\{1,2, \ldots ,N\}\setminus \al$ for the complement
of the set $\al$ and set $\ul n:= \{1,2,\ldots ,n\}$.
Let $|\al|$ denote the cardinality of the set $\al$.

For an arbitrary function $\Psi\in \bigwedge_1^N C_0^\infty(\bR^3)$, 
$\|\Psi\|=1$, and for $0\leq n\leq N$ we define 
$$
   \Gamma^n: 
= \Theta_{\ul n}\Bigg(\Tr_{\ul n^c} \Big[\Xi_{\ul n^c} |\Psi\rangle
 \langle \Psi|\Xi_{\ul n^c}\Big]\Bigg)\Theta_{\ul n} ,
$$
where $\Tr_{\ul n^c}$ denotes taking partial trace for the
$x_{n+1}, x_{n+2}, \ldots , x_N$ variables. Define
the fermionic Fock space as $\cF = \bigoplus_{n=0}^N \cH_n$
with $\cH_n:=\bigwedge^n\cH$ and we define a density matrix
$$
  \Gamma := \sum_{\al\subset \{ 1,2, \ldots ,N\}}\Gamma^{|\al|}=
\sum_{n=0}^N {N\choose n} \Gamma^n \quad \mbox{on} \;\;\cF.
$$

We first prove that $\Gamma\leq I$ on $\cF$. It is
sufficient to show that $\Gamma\leq I$ on the $n$-particle sectors 
for each $n$.
Let $n\leq N$, choose $\Phi\in\cH_n$,
and compute
\be
\begin{split}
\sum_{\al\subset\{1,2,
\ldots N\} \atop |\al|=n}& \langle \Phi,\Gamma^{|\al|}\Phi\rangle
=\sum_{\al\subset\{1,2,
\ldots N\} \atop |\al|=n} \int 
  \rd x_\al\rd x_\al' \ov \Phi(x_\al) \Gamma^n(x_\al, x_\al') \Phi(x_\al')\cr
  = & \sum_{\al} \int \rd x_\al \rd x_{\al}' \rd y_{\al^c}
 \ov \Phi(x_\al)
   \Theta_\al (x_\al)\Xi_{\al^c}(y_{\al^c})\Psi(x_\al, y_{\al^c})
  \ov\Psi(x_\al', y_{\al^c})\Xi_{\al^c}(y_{\al^c})\Theta_\al (x_\al')
\Phi(x_\al')\cr
\le & \sum_{\al} \int \rd x_\al \rd x_{\al}'\rd y_{\al^c}
\Theta_\al^2 (x_\al)\Xi_{\al^c}^2(y_{\al^c})|\Psi(x_\al, y_{\al^c})|^2
|\Phi(x_\al')|^2
\cr
= & \|\Phi\|_2^2\int \rd \bx |\Psi(\bx)|^2 \sum_\al\Theta_\al^2 (x_\al)
 \Xi_{\al^c}^2(x_{\al^c})\cr
 = & \|\Phi\|^2_2
\end{split}
\ee
using Schwarz inequality and that $1\equiv\prod_{j=1}^N
 [\theta^2(x_j) +\chi^2(x_j)]
= \sum_\al \Theta_\al^2(x_\al)\Xi^2_{\al^c}(x_{\al^c})$.

Second, for a fixed $n\leq N$, we compute
\be
\begin{split}
\Tr_\cF & \Big[ \Gamma\; \bigoplus_{n=0}^N \sum_{i=1}^n \fh_i\Big]
 = \sum_{n=0}^N  {N\choose n}
\Tr_\cF \Big[ \Gamma^n \sum_{i=1}^n \fh_i\Big]\cr
 & = 
 \sum_\al \sum_{i\in\al}
  \int \rd x_\al \rd x_{\al^c}\ov \Psi(x_\al, x_{\al^c})
  \Theta_\al(x_\al) \Xi_{\al^c}(x_{\al^c})
\fh_i\Big(
 \Theta_\al(x_\al) \Xi_{\al^c}   (x_{\al^c})\Psi(x_\al, x_{\al^c})
  \Big)\cr 
& = \sum_{i=1}^N \sum_{\al\; : \; i\in \al}  
\int \rd \bx \, \ov \Psi(x_\al, x_{\al^c})
  \Theta_{\al\setminus\{i\}}^2(x_{\al\setminus\{i\}})
 \Xi_{\al^c}^2(x_{\al^c})
\theta(x_i) \fh_i\Big( \theta(x_i)\Psi(x_\al, x_{\al^c})
  \Big)\cr
& = \sum_{i=1}^N \langle \Psi, \theta_i\fh_i\theta_i \Psi\rangle,
\end{split}
\ee
where the trace on the left hand side
is computed on $\cF$.
In the last
step we used that for any fixed $i$, we have
$1\equiv\prod_{j\neq i} [\theta^2(x_j) +\chi^2(x_j)]
= \sum_{\wh\al} \Theta_{\wh \al}^2\Xi^2_{\wh\al^c}$ where the
summation is over all $\wh\al\subset \{1, 2, \ldots, N\}\setminus \{i\}$
and $\wh\al^c =\{1, 2, \ldots, N\}\setminus \{i\}\setminus \al$.

A similar calculation for the two-body potential shows that
$$ 
\Tr_\cF \Big[ \Gamma \; \bigoplus_{n=0}^N\sum_{1\leq i<j\leq n} W_{ij}\Big] 
= \sum_{1\leq i<j\leq N}  \big\langle\Psi,
\theta_i\theta_j W_{ij} \theta_j\theta_i
\Psi\big\rangle\, .
$$
Thus, by the variational principle,
$$
  \Big\langle \Psi, \Big( \sum_{i=1}^N \theta_i\fh_i\theta_i
 + \sum_{1\leq i<j\leq N}
  \theta_i\theta_j W_{ij} \theta_j\theta_i \Big)\Psi\Big\rangle  \ge 
\inf_{\Gamma} \Tr_\cF \Bigg[ \Gamma\; \bigoplus_{n=0}^N
  \Big(\sum_{i=1}^n \fh_i +\sum_{1\leq i<j\leq n} W_{ij}\Big)\Bigg]\; .
$$
Since $\Gamma$ is a density matrix supported on $\Omega$, we obtain
\eqref{eq:abstr}. $\Box$

\end{document}